\documentclass[12pt]{article}
\usepackage{graphicx,a4wide,amsmath,amssymb,alltt,array}

\newcommand\ie{i.e.\ }
\newcommand\eg{e.g.\ }

\newcommand\uscore{\symbol{95}}

\newcommand\Code[1]{\ensuremath{\texttt{#1}}}
\newcommand\Var[1]{\ensuremath{\mathit{#1}}}
\newcommand\icon[1]{\lower 6pt\hbox{\includegraphics[scale=.8]{#124}}}

\makeatletter
\def\reportno#1{\gdef\@reportno{#1}}
\def\@maketitle{%
  \hfill{\small\begin{tabular}[t]{r}%
    \@reportno
  \end{tabular}\par}%
  \vskip 2em%
  \begin{center}%
    \let \footnote \thanks
    {\large \@title \par}%
    \vskip 1.5em%
    {
      \lineskip .5em%
      \begin{tabular}[t]{c}%
        \@author  
      \end{tabular}\par}%
    \vskip 1em%
    {
     \@date}%
  \end{center}%
  \par
  \vskip 1.5em}
\makeatother

\begin{document}

\reportno{MPP--2009--019\\arXiv:yymm.nnnn [hep-ph]}

\title{A Simple Way to Distribute Mathematica Evaluations}

\author{M. Bruhnke$^a$, T. Hahn$^b$ \\[.5ex]
$^a$Universit\"at W\"urzburg, \\
Am Hubland, D--97074 W\"urzburg, Germany \\[.5ex]
$^b$Max-Planck-Institut f\"ur Physik \\
F\"ohringer Ring 6, D--80805 Munich, Germany}

\date{February 11, 2009}

\maketitle

\begin{abstract}
We present a simple package for distributing evaluations of a 
Mathematica function for many arguments on a cluster of computers.
After setting up the hosts, the only change is to replace 
\Code{Map[f,\,points]} by \Code{MapCore[f,\,points]}.
\end{abstract}


\section{Introduction}

With the fairly recent arrival of low-cost multi-core CPUs, institutes 
often have significant computing power at their disposal.  Mathematica 7,
whose main motto is parallel computing, makes it relatively simple to send
a calculation to the fellow cores on the same machine, though still not 
exactly straightforward to distribute a calculation on a larger cluster.
The package we present in the following fills this gap.  After a one-time 
setup of the cluster, it allows to easily distribute calculations to as
many hosts as there are Mathematica licenses available (both ordinary 
licenses and Mathematica 7's sublicenses).

We certainly do not propose to parallelize `atomic' Mathematica
operations, like \Code{Simplify}, which is a daunting task even at the
conceptual level.  Rather, we focus on lengthy evaluations of one
function over many arguments, for example the evaluation of a
cross-section for many points in phase and/or parameter space. 
Incidentally, our package is not restricted to numerical evaluations,
but can handle any kind of Mathematica expressions.

Many physicists would argue that at least numerical evaluations of a
certain volume should be done in a compiled language for performance
reasons.  This is at best partially true, as Mathematica has a
formidable arsenal of functions, \eg for numerical analysis, which are
not easily available elsewhere, and it is the choice of algorithm that
influences the computation time much more than the speed of a single
evaluation.  Furthermore, in conjunction with MathLink, \eg through
FormCalc's Mathematica interface \cite{fcmma}, the execution speed is
essentially that of a compiled language and Mathematica's part is
`governing' the calculation.

The package we present in this paper is remarkably short and contains 
one main function \Code{MapCore} which substitutes \Code{Map} in 
serial calculations.  Sect.~\ref{sect:usage} describes usage of the 
package, Sect.~\ref{sect:ref} provides a function reference, and 
Sect.~\ref{sect:setup} describes installation and system setup.

\section{Overview}

\subsection{Usage of the MultiCore package}
\label{sect:usage}

The MultiCore package is loaded with
\begin{verbatim}
   << MultiCore`
\end{verbatim}
The next step is to add cores\footnote{%
	A note on nomenclature: we refer to a `core' as the 
	fundamental computation unit, \ie a processor able to run a 
	single thread.  A physical CPU may have several cores and 
	similarly a host may have several CPUs.}
on which evaluations can be distributed.  This can be done directly with 
\eg
\begin{verbatim}
   AddCore["pc123.mppmu.mpg.de"]
\end{verbatim}
or, if login under a different username is required,
\begin{verbatim}
   AddCore["batman@pc123.mppmu.mpg.de"]
\end{verbatim}
This explicit method becomes cumbersome, however, if many cores with 
varying loads are involved.  The alternate invocation
\begin{verbatim}
   AddCore[10]
\end{verbatim}
takes up to ten of the currently `free' cores.  This information is
supplied by the \Code{findcores} shell script (part of the MultiCore
package) which in turn reads the admissible cores from a
\Code{.submitrc} file and invokes \Code{ruptime} to determine the load.
The \Code{.submitrc} file has the simple syntax
\begin{verbatim}
   pc380   4
   pc381   4
   pc339b  2
   pc472
\end{verbatim}
where the optional integer behind the hostname indicates the number of
cores the host has.  The machines should be listed in descending CPU
speed, \ie fastest on top, to optimize performance.  Each remote host
should be running an \Code{rwhod} daemon, since then its load will be
reported through \Code{ruptime} and \Code{findcores} will use only the
free cores.

In the case of a Linux cluster, the \Code{.submitrc} file can be 
generated (more or less) automatically, with the help of the 
\Code{setupcores} script, as in:
\begin{verbatim}
   ./setupcores > $HOME/.submitrc
\end{verbatim}
This script assumes that the hosts are listed via \Code{ruptime}, that a 
password-free login via \Code{ssh} is possible, and that each host is 
running a flavour of Linux where \Code{/proc/cpuinfo} can meaningfully 
be read out.  The file generated in this way constitutes a `raw' version 
and should be reviewed by hand.

Each core launched requires a Mathematica license, \ie a Kernel license.
From Mathematica 7 on, each (main) license includes four sublicenses and
it is possible to use these sublicenses for parallelization (cf.\ Sect.\
\ref{sect:sublicense}, \Code{\$SublicenseFactor}).

One can further take care not to invoke more slave processes than
licenses available.  To this end \Code{AddCore} is invoked with an
integer $n\leqslant 0$, meaning that it should spawn at most so many
slaves that $|n|$ (main) licenses are left for other users.  Also one
can provide a second integer argument $m\leqslant 0$ to leave $|m|$
sublicenses unused.  This mode really makes sense only for network
licenses.  For non-network licenses, \Code{AddCore} silently assumes
that the other machines listed in \Code{.submitrc} have similar
licenses.

MultiCore generally works in a master--slave model, requiring one
license (but hardly any CPU time) for the master and one main or
sublicense for each slave.  We assume that all cores in the cluster run
the same Mathematica version, in particular that the master's version
number is the same as all slaves'.  In particular we assume that
subkernels on slave cores can be launched if and only if the master is
running Mathematica 7.

Quitting the master's Mathematica Kernel automatically closes all 
links, so explicitly `removing' registered cores is usually not 
necessary unless one wants to free Mathematica licenses.  Each slave 
session is characterized by an identifier of the form 
\Code{host[id]}, where \Code{host} is the host name and \Code{id} an 
integer link id.  The syntax for \Code{RemoveCore} is 
\begin{verbatim}
   RemoveCore[host]
   RemoveCore[host[id]]
\end{verbatim}
where both \Code{host} and \Code{id} may be a pattern.  Thus, 
\Code{RemoveCore["pc123"]} closes all slaves on host \Code{pc123} and
\Code{RemoveCore[\uscore]} closes all current slave sessions.

Once the cores are registered, the only necessary substitution is to 
replace \Code{Map} (\Code{/@}) by \Code{MapCore} to make multiple 
evaluations execute in parallel.

\textbf{Important:} The only slightly non-straightforward aspect is the
remote definition of the function being evaluated.  \Code{MapCore}
sends the definition of this function to the slave as much as the
\Code{Save} function would save it in a file.  This \emph{fails to work}
(for both \Code{MapCore} and \Code{Save}) if the function depends on
a \Code{LinkObject} in the master's session, \ie if the function is or
invokes a MathLink function.  Even if the slave session has the same
MathLink executable installed, it will in general not communicate via
the syntactically same \Code{LinkObject}.

To work around such cases, the \Code{AddCore} function has an optional 
second argument.  This argument is sent to the slave upon opening of the 
link as an initialization command.
In our opinion the best procedure in the MathLink case mentioned above 
is not to install the MathLink executable in the master's session at 
all, to prevent sending any explicit \Code{LinkObject} pointing to the 
master's installed MathLink executables, and instead include the 
\Code{Install} statement in the \Code{AddCore} invocation, as in
\begin{verbatim}
   AddCore[0, Install["LoopTools"]]
\end{verbatim}
Also, if the function has a very lengthy definition one might want to 
place it in a file and load that via the initialization command, \eg
\begin{verbatim}
   AddCore[0, << myfunction.m]
\end{verbatim}
Of course one would have to submit this file to each slave first if they
do not have access to the master's filesystem.  Note, however, that the
slaves' working directory is the user's home directory, not the current
working directory on the master.  In other words, the file to be loaded
must include a path unless it resides in the home directory anyway.

\subsection{MultiCore's concurrency handling}
\label{sect:concurrency}

\Code{MapCore} tries to have the given points calculated as quickly as
possible.  Therefore it distributes more (less) than \Var{patchsize}
points to faster (slower) cores by evaluating its internal timing
statistics.  Once all points of the list are distributed, \Code{MapCore}
redistributes the unfinished points until the result for all points are
available.  It automatically decreases the patchsize according to the
remaining list size, too.  Although due to the competition $N-1$ cores
are not yet finished when \Code{MapCore} returns, the time until all
slaves are again ready is negligible.  The identifier \Code{\$CallID}
helps \Code{MapCore} to distinguish between new and old data of multiply
distributed points.

\subsection{MultiCore's error handling}

Especially during long parallized calculations of many
CPU-time-expensive points, link error handling plays an important role.
If the link to one host, \ie one or more cores, is lost, \Code{MapCore}
redistributes the as yet uncalculated points to the remaining hosts,
executes the equivalent \Code{RemoveCore} call and prints a warning
message.  After \Code{MapCore} has returned one might want to add the
lost host by re-invoking \Code{AddHost}.

\section{MultiCore package Function Reference}
\label{sect:ref}

\subsection{AddCore}
\label{sect:addcore}

\Code{AddCore} adds (registers) cores, \ie opens links to remote 
machines for subsequent distributed evaluation with \Code{MapCore}.  It 
is invoked in one of the following ways:
\begin{itemize}
\item \Code{AddCore[\Var{hostname}]} adds one core on \Var{hostname} 
using a main license.

\item \Code{AddCore[\Var{hostname},\,"subkernel"]} adds one core on
\Var{hostname} using a sublicense.

\item \Code{AddCore[\Var{n}]} ($\Var{n} > 0$, integer) adds up to
\Var{n} cores using the \Code{findcores} script (described below) 
using a ratio \Code{\$SublicenseFactor} : 1 of sublicenses to main
licenses (cf.\ Sect.\ \ref{sect:sublicense}).

\item \Code{AddCore[\Var{n}]} ($\Var{n} \leqslant 0$, integer) adds as 
many cores as there are main licenses using \Code{findcores}, but 
leaves at least $|\Var{n}|$ main licenses for other users.

\item \Code{AddCore[\Var{n},\,\Var{m}]} ($\Var{n}, \Var{m}$ integer)
same as above, with \Var{n} for main licenses and \Var{m} for
sublicenses.

\end{itemize}

The last two invocations really make sense only for network licenses. 
For non-network licenses, it is silently assumed that the information
taken from \Code{\$LicenseProcesses} and \Code{\$MaxLicenseProcesses}
(in the master's session) holds also for the remote cores.  Each link
corresponds to one core on a remote machine.  It is hence permissible to
add the same host more than once, to account for its number of cores. 
The links are identified, apart from the hostname, by a unique integer
link id.  This id is also sent to each slave process as \Code{\$CoreID}
and can be used to \eg construct unique filenames. Core additions are
cumulative.  Links are released either through explicit removal with
\Code{RemoveCore} or by quitting the master's Mathematica Kernel.

The \Code{findcores} script is part of the MultiCore package.  It needs a 
\Code{.submitrc} file in which the admissible cores for distributed 
computing are listed.  Each line has the syntax
\begin{verbatim}
   hostname   [# of cores]
\end{verbatim}
Comment lines starting with a \Code{\#} are allowed.  Cores are 
processed in sequential order, \ie the fastest machine should appear at 
the top of this list.  The \Code{.submitrc} file is searched for in the 
following order:
\begin{itemize}
\item \Code{./.submitrc},
\item \Code{\$HOME/.submitrc},
\item \Code{\Var{(MultiCore\,installation\,directory)}/submitrc},
\item \Code{/usr/local/share/submitrc}.
\end{itemize}
\Code{findcores} invokes \Code{ruptime} to determine the load on a
remote machine.  This works only if the remote machine is running an 
\Code{rwhod} daemon.  If not, the load is assumed to be zero, \ie all 
cores are taken.

\subsection{RemoveCore}

\Code{RemoveCore} removes (unregisters) cores from the internal list,
shuts down the corresponding remote kernels and closes the links.  Each
core is identified by two quantities, the hostname and the link id. 
Calling \Code{RemoveCore} is usually not necessary, as quitting the
master's Mathematica Kernel automatically closes all links.
\begin{itemize}
\item \Code{RemoveCore[\Var{hostname}[\Var{id}]]} removes all cores
matching \Var{hostname} and \Var{id}, where either may contain a
pattern.  For example, \Code{RemoveCore[\uscore]} removes all links, and
\Code{RemoveCore["pc456"[\uscore]]} removes all links to \Code{pc456}.

\item \Code{RemoveCore[\Var{hostname}]} is equivalent to 
\Code{RemoveCore[\Var{hostname}[\uscore]]}.
\end{itemize}

\subsection{ListCore}

\Code{ListCore} lists the currently registered cores.
\begin{itemize}
\item \Code{ListCore[\Var{hostname}[\Var{id}]]} lists all cores
matching \Var{hostname} and \Var{id}, where either may contain a 
pattern.  \Code{ListCore[\uscore]} thus lists all cores.

\item \Code{ListCore[\Var{hostname}]} is equivalent to 
\Code{ListCore[\Var{hostname}[\uscore]]}.
\end{itemize}

\subsection{MapCore}

\Code{MapCore} is the main function of the MultiCore package.  It 
substitutes \Code{Map} in serial calculations.
\begin{itemize}
\item \Code{MapCore[\Var{f},\,\Var{points},\,\Var{patchsize}]}
distributes the computation of \Var{f} for all items in \Var{points} to
the cores previously registered with \Code{AddCore}.
\end{itemize}

The integer argument \Var{patchsize} is optional (default value: 5) and
tells \Code{MapCore} how many points on average should be sent to each
core.  As every set of results returned by a slave contains timing
information, the master distributes points according to the slaves'
performance.  Until the master has gathered enough statistics about the
slaves' timings it sends exactly \Var{patchsize} points to each core.

The larger the computation time for a single point is, the smaller
\Var{patchsize} should be chosen.  A smaller value may also be
profitable if the participating cores have significant differences in
speed.  A \Var{patchsize} of 1 achieves the best load-levelling but
incurs the highest communication overhead.  We have generally found the
communication overhead to be negligible if the computation time for one
patch is several seconds or more (see also performance tests in Section
\ref{sect:performance}).

\subsection{RemoteMath}
\label{sect:remotemath}

\Code{RemoteMath} encodes the invocation of a remote Mathematica Kernel.  
It receives one arguments and one flag, the hostname and the type of 
license which shall be used while launching the kernel. If required
one can define different invocation strings for different hosts. 

\begin{itemize}
\item \Code{RemoteMath[\Var{host},\,\Var{opt}] := \Var{remotestring}} 
defines \Var{remotestring} as the command for invoking a remote
Mathematica Kernel on \Var{host}.  Options for the remote kernel are
given in \Var{opt}, which is presently restricted to \Code{-subkernel}
for launching a subkernel.
\end{itemize}

The default command is
\begin{verbatim}
ssh (host) 'exec /bin/sh -lc \
  "test `uname -s` = Darwin && nice -19 MathKernel (opt) -mathlink \
                            || nice -19 math (opt) -mathlink"'
\end{verbatim}
This is an \Code{ssh} command which starts a remote login shell that 
executes, with nice 19, \Code{MathKernel} on MacOS and \Code{math} on 
other systems.  Starting a login shell is important as it sources the 
shell's initialization files, which may modify the PATH.

If the Mathematica Kernel executable cannot be started using this
command because it is not on the PATH, we recommend adding the
appropriate directories to the PATH on the remote system rather than
modifying the \Code{RemoteMath} definition.

\subsection{RemoteMap}

With \Code{RemoteMap} one can specify a mapping function which shall be
applied on all remote hosts, \ie slave sessions, to the point patches
they receive from the master.  Its default
\begin{verbatim}
RemoteMap[f_, points_] := Map[f, points]
\end{verbatim}
is the usual \Code{Map} function.  This may be overwritten with an
individual function which must have the same argument structure as
\Code{Map[\Var{f},\,\Var{points}]}.  This feature could for example be
used to leave a part of the parallelization to Mathematica 7 using the
\Code{ParallelMap} function.  In that case one of course would set the
number of cores in \Code{.submitrc} to 1 for all hosts.

\subsection{\$FindCores}

\Code{\$FindCores} contains the full path to the \Code{findcores} script,
including (if necessary) any options.  The full syntax of \Code{findcores}
is:
\begin{verbatim}
   findcores [-f rcfile] [-h ruptimehost]
\end{verbatim}
where \Code{rcfile} specifies the explicit location of the \Code{submitrc}
file (see Sect.~\ref{sect:addcore}) and \Code{ruptimehost} specifies the
host on which to invoke \Code{ruptime} to find out the load of the
machines listed in the \Code{submitrc} file.  The latter is necessary if 
running the master process on a machine not connected to the cluster, \eg
a laptop.

Note: changing \Code{\$FindCores} modifies subsequent invocations of
\Code{AddCore} only, \ie links once established are not changed by a
different value of \Code{\$FindCores}.

\subsection{\$MsgLevel}

\Code{\$MsgLevel} specifies how verbose the master--slave communication 
is reported on screen.
\begin{itemize}
\item \Code{\$MsgLevel = \Var{n}} sets the message level to \Var{n}.

The default message level is 1, which just reports the adding and 
removing of cores as well as link failures.
\end{itemize}

\subsection{\$CoreID}

\Code{\$CoreID} is unique identifier for each slave session.
\begin{itemize}
\item \Code{\$CoreID} (in the master's session) is the id of the last 
slave session spawned.  This number should not be tampered with.

\item \Code{\$CoreID} (in the slave's session) is a unique identifier
of the session.
\end{itemize}

\subsection{\$CallID}

\Code{\$CallID} is available in both the master and slave session.
In the master session it counts the total number of calls to
\Code{MapCore}.  In the slave session it identifies that certain call
to \Code{MapCore} which invoked the last computation on this slave.
Note that they do not have to be equivalent (see Sect.\ 
\ref{sect:concurrency}).

\subsection{\$SublicenseFactor}
\label{sect:sublicense}

The integer \Code{\$SublicenseFactor} is a global parameter in the
master session which is set to 4 if the Mathematica version is 7 or
above, and 0 otherwise.  Only \Code{AddCore[\Var{n}]} with $n > 0$ makes
use of it to decide how many sublicenses should be used launching a
kernel before using another (rare) main license.  Setting
\Code{\$SublicenseFactor} manually only makes sense if one uses
Mathematica 7 and wants to optimize it to the mean ratio of unused
sublicenses to unused main licenses which might be greater than 4 in
some cluster networks.

\subsection{\$ListPositions}
\label{sect:listpositions}

\Code{\$ListPositions} is available in the slave session only.
This list contains the positions of the points in the original list 
which are to be evaluated by the slave.

Both \Code{\$CallID} and \Code{\$ListPositions} can \eg be used to 
construct unique filenames.  For example, if a single evaluation is very 
costly in CPU time, one may want to store each result immediately after 
computation.  This could be solved through a wrapper function
\begin{verbatim}
RemoteMap[f_, points_] :=
  MapThread[store[f], {points, $ListPositions}]

store[f_, dir_:"results"][x_, i_] :=
Block[ {file = ToFileName[dir, ToString[i]]},
  If[ FileType[file] === File,
    Get[file],
  (* else *)
    If[ FileType[dir] === None, CreateDirectory[dir] ];
    (Put[#, file]; #)& @ f[x] ]
]
\end{verbatim}
Results for each point would be stored in \Code{results/\Var{n}}, where 
\Var{n} is each point's index in the original list.
In addition to \Code{\$ListPositions} one could use \Code{\$CallID} to
generate unique filenames over multiple invokations of \Code{MapCore} in 
the same master session.

\section{Performance Tests}
\label{sect:performance}

We tested the performance and scalability properties of MultiCore on
both a homogeneous and inhomogeneous cluster of 25 cores for different
evaluation times per point (tpp) and different patchsizes. As a testing
function we used a simple pause directive
\begin{verbatim}
  f[p_][x_] := (Pause[p]; x)
\end{verbatim}
and mapped it over 10000 resp.\ 1000 arbitrary points for different
numbers of cores ranging from 0 (local evaluation), 1 (slave) to 25
(slaves) and pausing times $p = 0.01, 0.1, 1$ seconds.

\begin{figure}
\begin{tabular}{cc}
\includegraphics[width=.5\hsize]{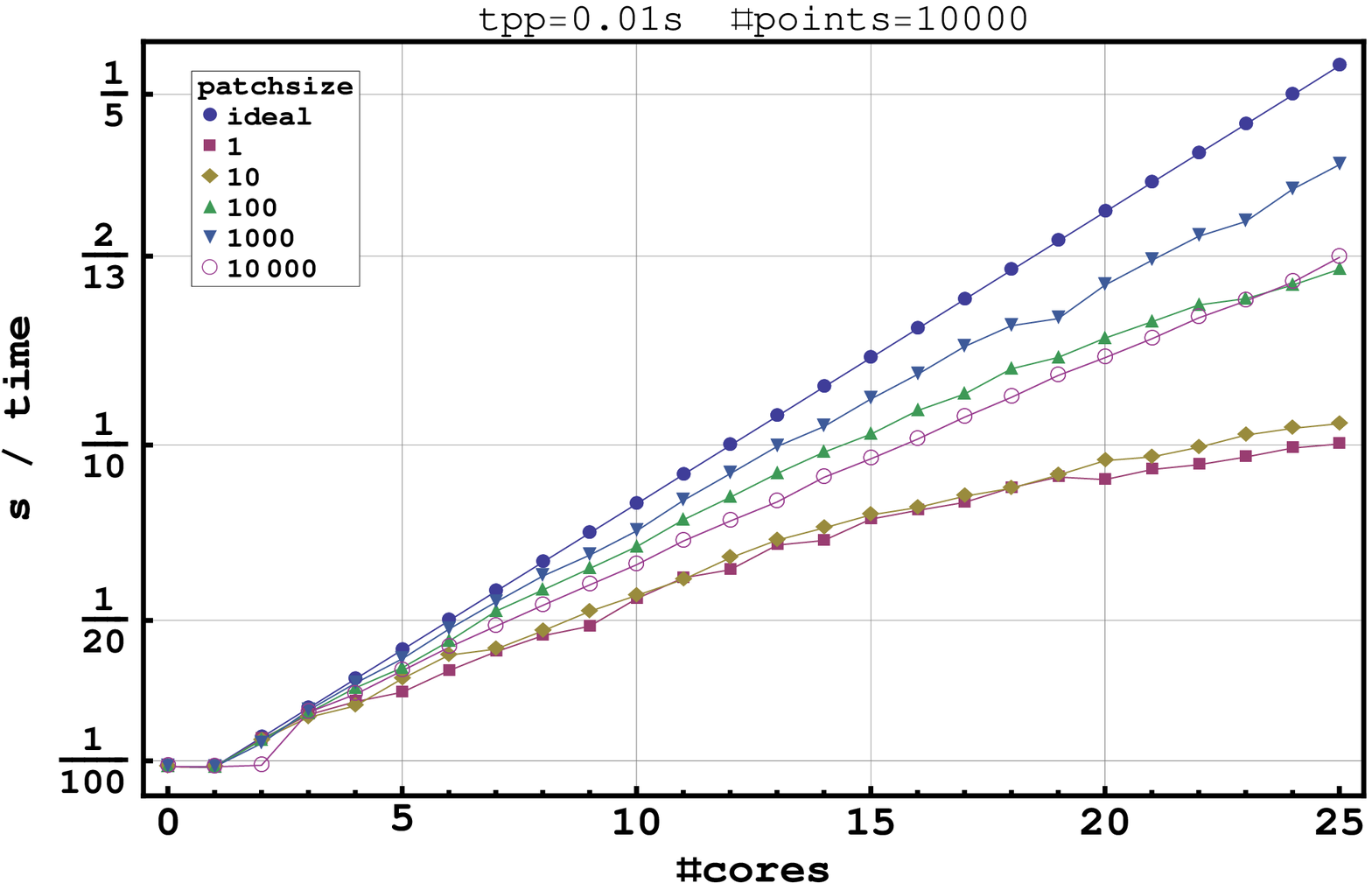} &
\includegraphics[width=.5\hsize]{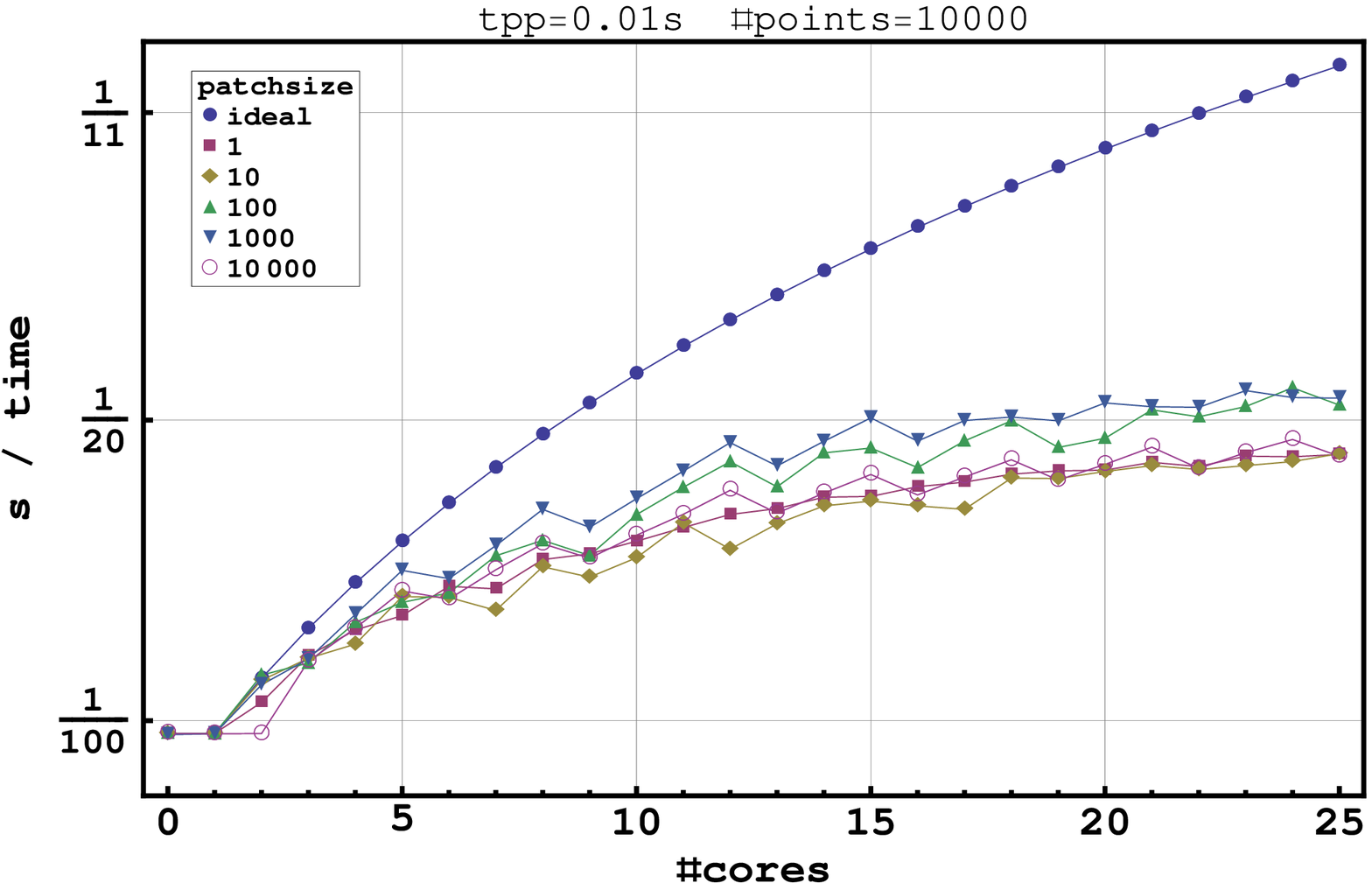} \\
\includegraphics[width=.5\hsize]{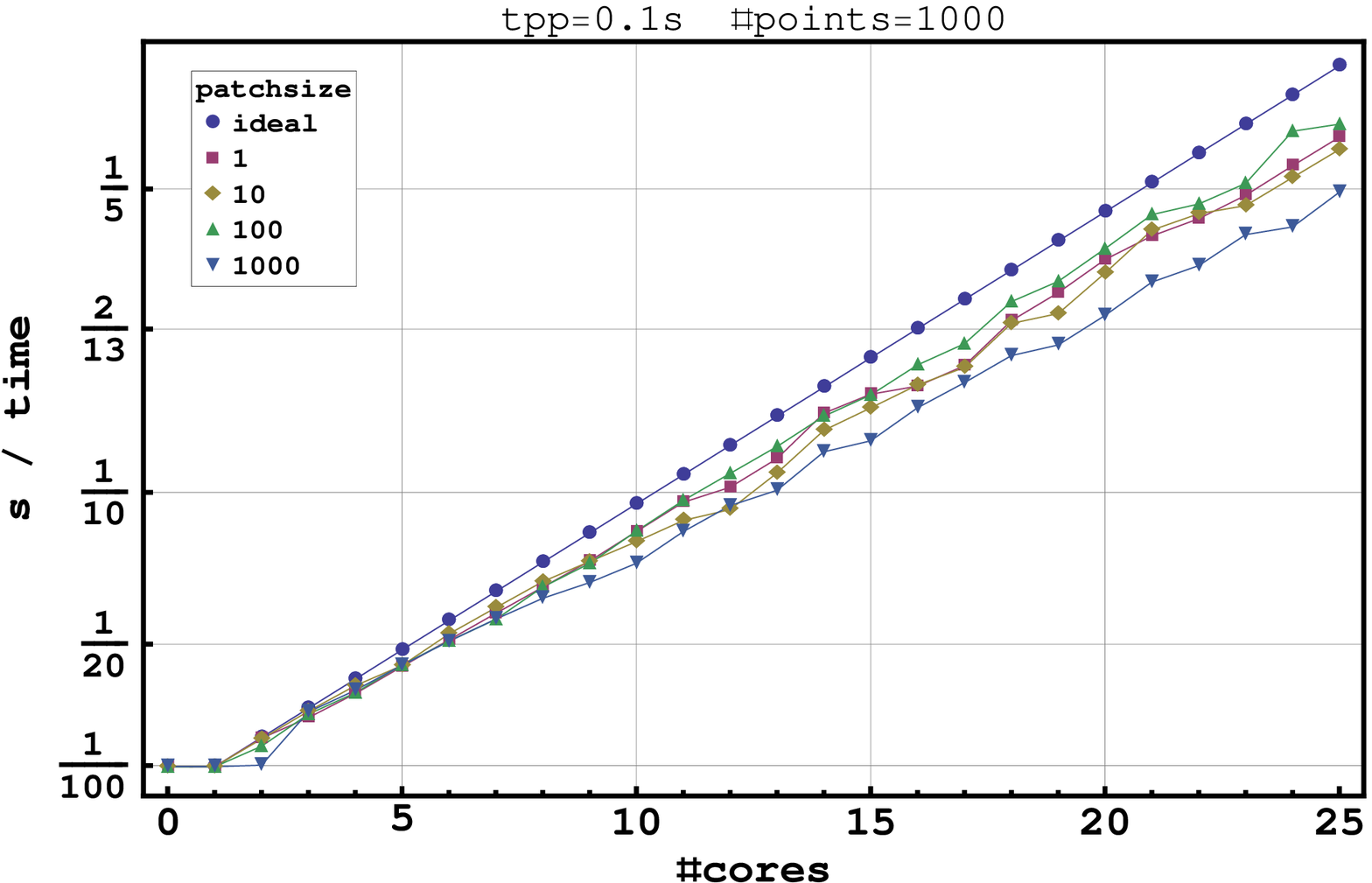} &
\includegraphics[width=.5\hsize]{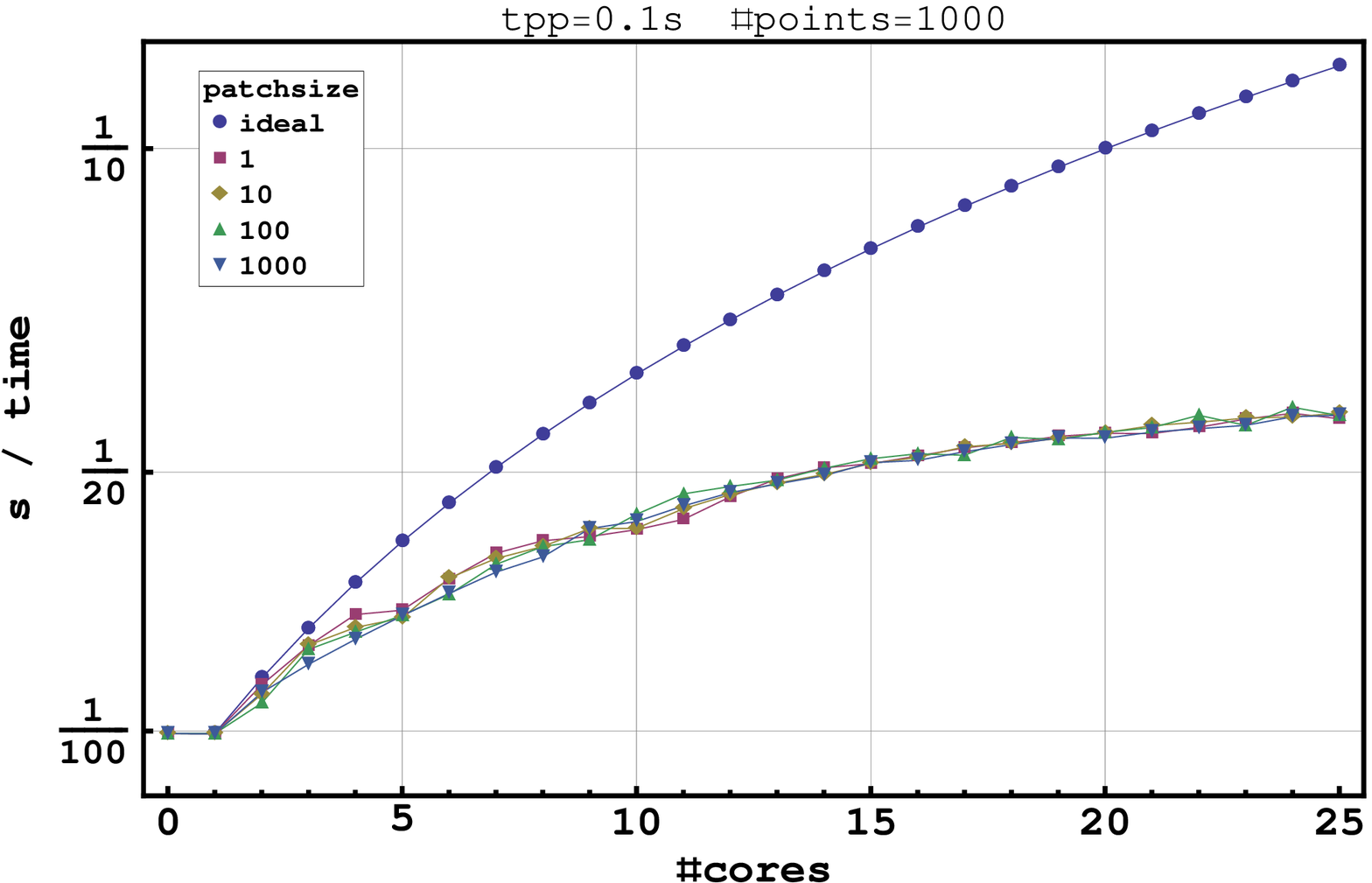} \\
\includegraphics[width=.5\hsize]{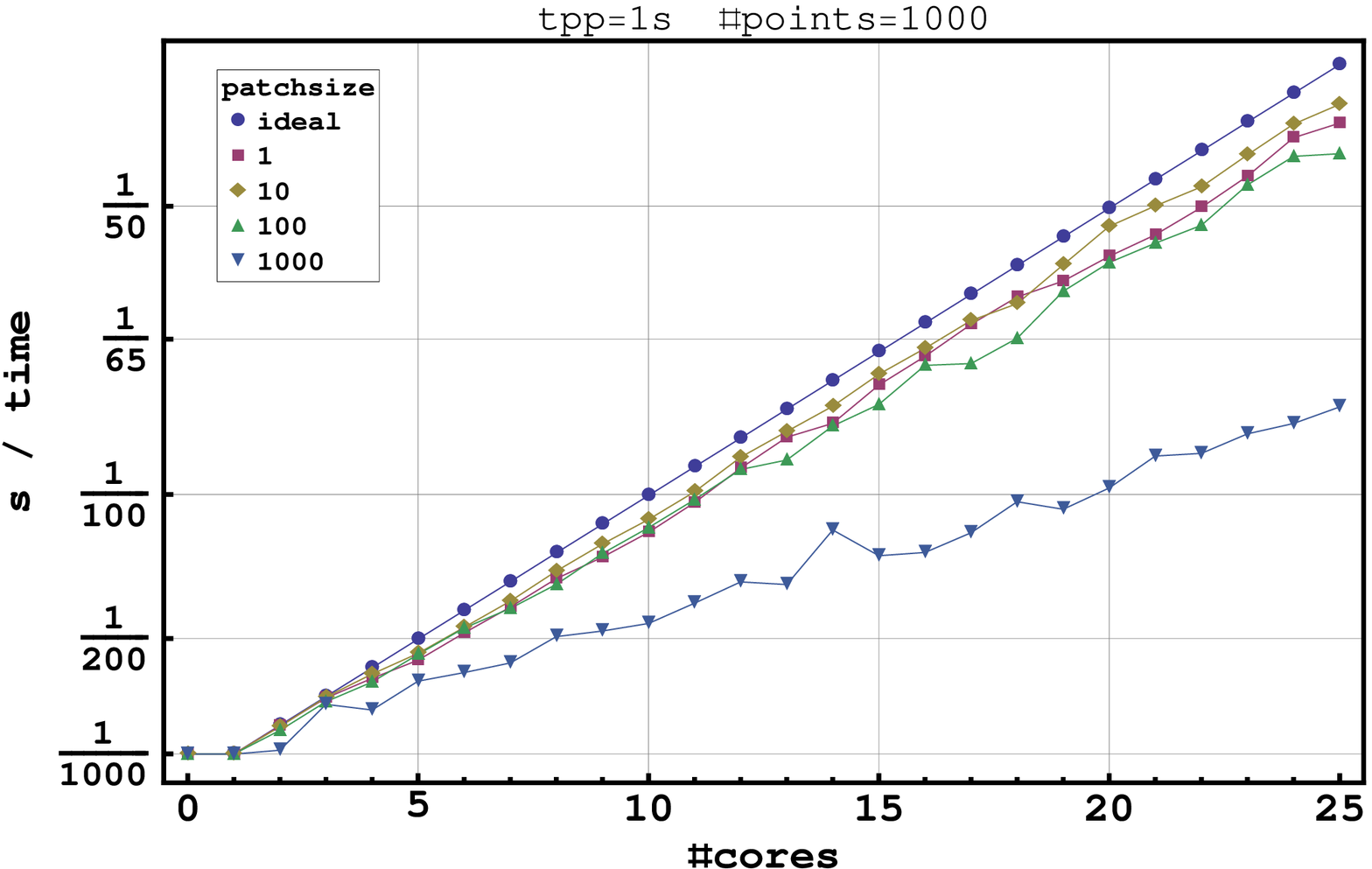} &
\includegraphics[width=.5\hsize]{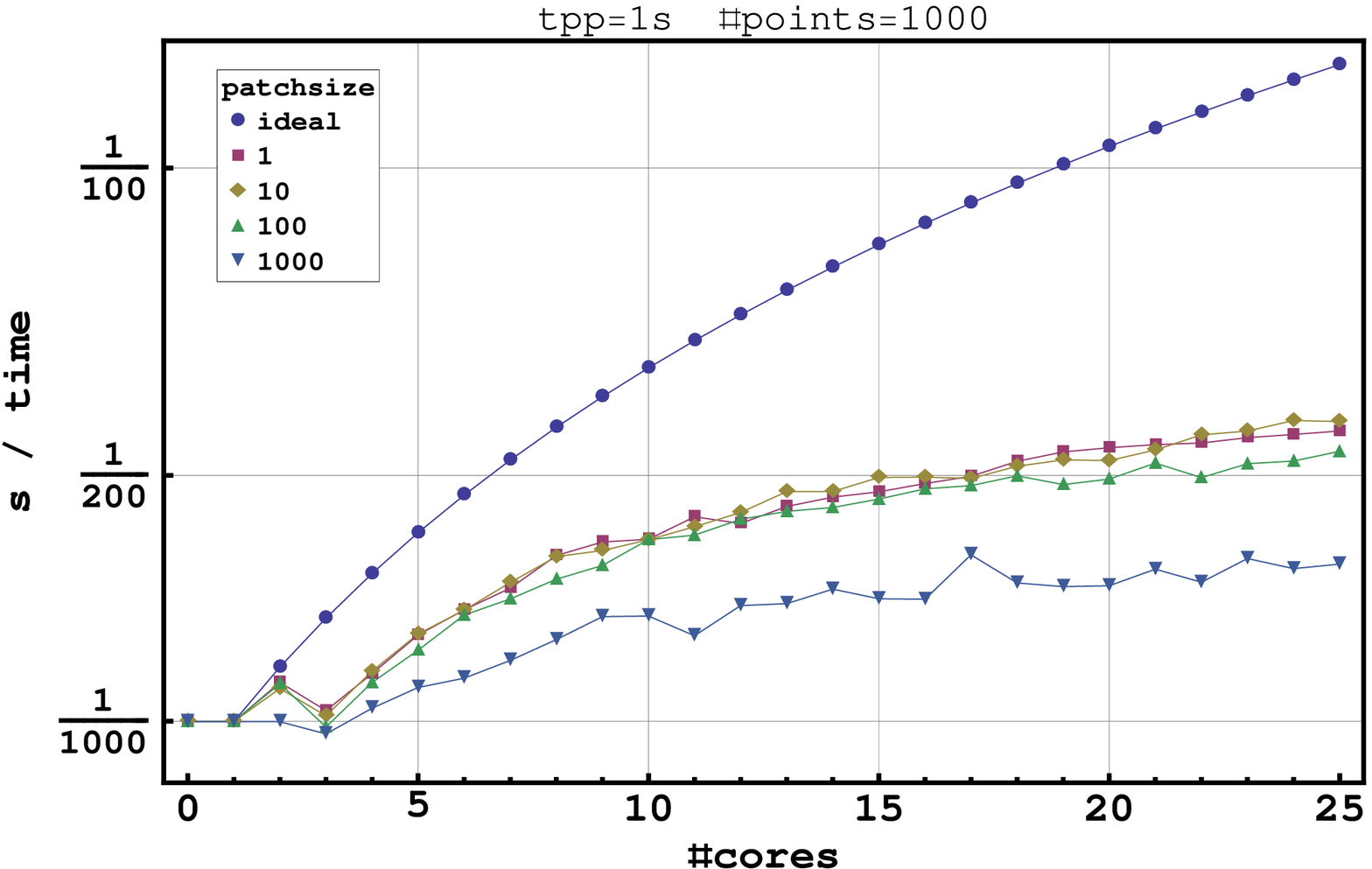}
\end{tabular}
\vspace*{0ex}
\caption{\label{fig:perf}Reciprocal total timings as a function of 
number of cores for different evaluation times per point (tpp), 
different number of points (see heading of corresponding plot) and
patchsizes. The left column shows the result for the homogeneous 
cluster ($\text{tpp}_i = \text{tpp}_1 = \text{const}$). The right 
column shows those for the inhomogeneous cluster \ie 
$\text{tpp}_i = \text{tpp}_1 (1 + 3\,\frac{i-1}{24})$ for 
$i=1,\ldots,25$.} 
\end{figure}

In case of the homogeneous cluster we assumed a constant `evaluation'
time per point for each core.  In the ideal case one would expect the
total time to be inversely proportional to the number of cores.  The
three inverse plots on the left-hand side of Figure \ref{fig:perf}
illustrate this ideal connection (blue line) and the deviation of the
measured timings for different patchsizes.  We took the average of ten
independent runs for each point.  As one can see, MultiCore's
performance in a homogeneous cluster barely depends on the (reasonable
choosen) patchsize.  It shows an almost perfect scaling behaviour for
evaluation timings per point of around 0.1 seconds or more.  For smaller
tpp's one would better choose a bigger patchsize.

To simulate an inhomogeneous cluster we linearly spread the tpp's from
\eg 1.0 to 4.0 seconds over the range of the 25 cores.  On a subset of
\eg 14 cores we of course added the 14 fastest ones.  Due to the
different evaluation timings the ideal curve is no longer a line. 
Instead, in the ideal case the total time $T$ depends on the number and
performance of the added cores:
$$
\frac 1T = \frac 1n \sum_{i=1}^{N} \frac 1{\text{tpp}_i}
$$
with $\text{tpp}_i$ being the tpp of core $i$ and $N$ the number of
cores and $n$ the total number of points.  The three plots on the right
hand side of Figure \ref{fig:perf} show the testing results for
different tpp's (of the fastest core) and for different patchsizes.
Again, the patchsize is not a crucial parameter.  As before, deviations
occur for the small tpp = 0.01 sec.  The scaling behaviour for large
numbers of cores seems to be at most satisfactory since MultiCore's
parallalizing takes about twice as long as the ideal case predicts. 
But if one compares the total timings of 25 unequal cores to the
corresponding timings on the left hand side, one sees that it takes only
about 10 cores from the homogeneous cluster to do the same job. 
Therefore one principally has to consider the performance gain before
joining much slower cores to one's cluster.

\section{Installation and System Setup}
\label{sect:setup}

The MultiCore package is available from
\Code{http://www.feynarts.de/multicore}.  Installation is as simple as
unpacking the tar file.  MultiCore requires Mathematica versions 5 and up
(version 7 preferred).

To be able to load MultiCore regardless of the current directory, the
MultiCore installation directory has to be added to Mathematica's
\Code{\$Path}, for example by placing a statement like
\begin{verbatim}
   PrependTo[$Path, "/my/path/to/MultiCore"]
\end{verbatim}
in \Code{\Var{prefdir}/Kernel/init.m}, where \Var{prefdir} is one of
\begin{itemize}
\item \Code{/usr/share/Mathematica}
      (system-wide, Linux),
\item \Code{\$HOME/.Mathematica}
      (user-specific, Linux),
\item \Code{/Library/Mathematica}
      (system-wide, MacOS),
\item \Code{\$HOME/Library/Mathematica}
      (user-specific, MacOS),
\item \Code{\$ALLUSERSPROFILE/Application Data/Mathematica} 
      (system-wide, Cygwin),
\item \Code{\$USERPROFILE/Application Data/Mathematica}
      (user-specific, Cygwin).
\end{itemize}

The package has been tested under Linux, MacOS, and Windows/Cygwin, both
as master and as slave.  The communication with remote Mathematica
Kernels requires attention to a few details that may not be obvious:
\begin{itemize}
\item An \Code{sshd} daemon must be running on the remote machine and
access not restricted by a firewall.  On Cygwin one has to start
\Code{sshd} once with ``\Code{net start sshd}'' (as Administrator) and
on MacOS one has to open the ssh port in the firewall (System
Preferences -- Sharing -- Remote Login).

\item ssh access to remote machines must be possible without password
authentication.  This requires that a host key is generated with
\Code{ssh-keygen} and the public part of it (typically
\Code{\$HOME/.ssh/id\uscore rsa.pub}) copied to
\Code{\$HOME/.ssh/authorized\uscore keys}.

\item If remote access other than by ssh is required, one needs to
redefine the \Code{RemoteMath} function, which encodes the command
string used to execute remote Mathematica Kernels (see
Sect.~\ref{sect:remotemath}).  This can either be done in the master
session before any \Code{AddCore} invocations, or once and forever in
\Code{MultiCore.m}.
\end{itemize}

\section{Summary}

The MultiCore package provides a simple mechanism to distribute
(parallelize) evaluations of a single functions over many points.  After
setting up the cores participating in the calculation with
\Code{AddCore}, the single replacement of \Code{Map} by \Code{MapCore}
suffices to distribute the calculation. \Code{MapCore} is not limited to
numerical evaluations, but can handle any type of Mathematica
expression.

From Mathematica 7 on, parallelization on several cores of a single host
is a built-in functionality.  Distributing calculations over more than
one host is not straightforward, however, but can be done with the same
ease using the \Code{MultiCore} package.

The package is open source and is licensed under the GPL.  It can be
downloaded from \Code{http://www.feynarts.de/multicore} and runs on
Mathematica versions 5 and up (version 7 recommended).

\section*{Acknowledgements}

We thank A.~Hoang for playing our guinea pig in the beta stage
and apologize to the MPI users for using up too many Mathematica 
licenses during testing.

\end{document}